# Can narrow disks in the inner solar system explain the four terrestrial planets?

Patryk Sofia Lykawka[1]

[1] School of Interdisciplinary Social and Human Sciences, Kindai University, Shinkamikosaka 228-3, Higashiosaka, Osaka, 577-0813, Japan; patryksan@gmail.com

## ABSTRACT

A successful solar system model must reproduce the four terrestrial planets. Here, we focus on 1) the likelihood of forming Mercury and the four terrestrial planets in the same system (a 4-P system); 2) the orbital properties *and* masses of each terrestrial planet; and 3) the timing of Earth's last giant impact and the mass accreted by our planet thereafter. Addressing these constraints, we performed 450 N-body simulations of terrestrial planet formation based on narrow protoplanetary disks with mass confined to 0.7–1.0 au. We identified 164 analogue systems, but only 24 systems contained Mercury analogues, and eight systems were 4-P ones. We found that narrow disks containing a small number of embryos with individual masses comparable to that of Mars and the giant planets on their current orbits yielded the best prospects for satisfying those constraints. However, serious shortcomings remain. The formation of Mercury analogues and 4-P systems was too inefficient (5% and 2%, respectively), and most Venus-to-Earth analogue mass ratios were incorrect. Mercury and Venus analogues also formed too close to each other (~0.15–0.21 au) compared to reality (0.34 au). Similarly, the mutual distances between the Venus and Earth analogues were greater than those observed (0.34 *vs*. 0.28 au). Furthermore, the Venus–Earth pair was not reproduced in orbital-mass space statistically. Overall, our results suggest serious problems with using narrow disks to explain the inner solar system. In particular, the formation of Mercury remains an outstanding problem for terrestrial planet formation models.

# 1. INTRODUCTION

The terrestrial planets formed via the accretion of embryos and planetesimals in a protoplanetary disk less than a few hundred Myr after the birth of the solar system (Morbidelli et al. 2012). Embryo and planetesimal formation are active fields of research that predict various possible outcomes for the disk (e.g. see reviews by Morbidelli & Raymond 2016 and Raymond et al. 2020 for more details), but a typical outcome can be summarised as follows. After the dispersal of gas from a nebular disk early in the life of the solar system, embryos existed with masses varying from those of dwarf planets to that of Mars, as did residual small planetesimals in a disk that extended to the asteroid belt (Kokubo & Ida 2002; Morishima et al. 2010; Chambers 2016; Walsh & Levison 2019). Based on such initial protoplanetary disk conditions, early studies on terrestrial planet formation managed to produce planets with masses comparable to those of Venus and Earth (Chambers 2001; Raymond et al. 2009; Lykawka & Ito 2013; Fischer & Ciesla 2014), but they generally failed to reproduce the small mass of Mars. To circumvent this problem, recent terrestrial planet formation models claim that the disk embryos or planetesimals were mostly or even completely concentrated in a narrow annulus with edges that roughly correspond to the current semi-major axes of Venus and Earth. (Hansen 2009; Walsh et al. 2011, Walsh & Levison 2016). In this way, these models imply that Mars acquired its small mass as a result of the lack of disk mass beyond ~1 au. If this is true, Mars formed inside such narrow disks and was later scattered outward to its current orbit at a semi-major axis $a \approx 1.5$ au. Alternatively, Mars may have formed in a mass-depleted region of the disk near its current orbit[1].

By contrast to earlier works on terrestrial planet formation, Hansen (2009; henceforth H09) presented a strong case in favour of narrow disks. Because H09 considered disks with their masses entirely confined within 0.7–1.0 au, they found that more massive planets formed around the same region, and less massive planets were scattered out to adjacent regions, less than 0.7 au or greater than 1.0 au. In this way, the obtained systems in those disks resembled the currently observed dichotomy of large-mass planets (Venus and Earth) being surrounded by small-mass planets (Mercury and Mars). H09 considered 38 simulation runs of gas-free protoplanetary disks containing 400 embryos distributed randomly within a disk with initial eccentricity $e = 0$ (circular orbits) and inclination $i = 0.5°$. Only Jupiter was included in their simulations, set with an orbit similar to its current one, $a = 5.2$ au and $e = 0.05$. Finally, their systems were evolved until 1 Gyr.

The results of H09 inspired several subsequent works. Perhaps the most influential was the so-called Grand Tack model (Walsh et al. 2011; Jacobson & Morbidelli 2014; Brasser et al. 2016a; Walsh & Levison 2016), which proposed that a gas-driven migration of Jupiter could truncate the initially extended disks at the end of gas dispersal. Similar to the narrow disk postulated in H09, such truncated disks[2] possess mass concentrated within ~1–1.5 au. Thus, the Grand Tack model can also form Mars-like planets with masses comparable to that of Mars. Clement et al. (2019a) found similar results after replicating the narrow disk of H09 and testing the influence of the dynamical instability of giant planets on that kind of disk. Other studies investigated the disk of H09 by including the effects of disk gas and fragmentation (Walsh & Levison 2016; Deienno et al. 2019). These studies found that Mars-like embryos formed within one to a few Myr and that the final systems had similar properties to those found by the Grand Tack and H09 models. Raymond & Izidoro (2017) replicated the H09 disk to investigate formation of the asteroid belt, but they also obtained small-mass Mars-like planets as by-products. In sum, the findings of H09 serve as the basis for several models of the formation of Mars.

---

[1] Currently, there are five models for the formation of Mars: Grand Tack (Walsh et al. 2011), empty asteroid belt (Raymond & Izidoro 2017), early instability (Clement et al. 2018), pebble accretion (Levison et al. 2015), and sweeping secular resonance (Bromley & Kenyon 2017). Please refer to Raymond et al. (2020) and Lykawka & Ito (2019) for detailed discussions and criticisms of these models.

However, note that the initial conditions and implications of truncated and narrow disks differ somewhat in the Grand Tack and H09 models. For instance, truncated disks typically possess a mass-depleted component beyond ~1–1.5 au that is not considered in the narrow H09 disk. This component can strongly affect the formation of Mars and the asteroid belt, and can deliver volatiles to the forming terrestrial planets (Izidoro et al. 2014, 2015; O'Brien et al. 2014; Lykawka & Ito 2019). Although the formation of Mercury is addressed in H09, this discussion is omitted in representative work with the Grand Tack model. Furthermore, according to these models, truncated disks typically contain 10–80 embryos within ~1–1.5 au, whereas narrow disks have 400 embryos confined at 0.7–1.0 au at the time that disk gas is dispersed. Conversely, embryos in truncated disks are also more massive (typically 0.025–0.1 $M_\oplus$ vs. 0.005 $M_\oplus$ in narrow disks). Although truncated or narrow disks are often considered an essential initial condition for success in forming a small-mass Mars-like planet, there has been little discussion of the role that any of these distinct disk properties might play during terrestrial planet formation.

Lykawka & Ito (2019; henceforth LI19) investigated the properties of protoplanetary disks that could form three- or four-planet analogue systems containing Mercury or Mars analogues in addition to the Venus–Earth analogue pair in the same system[3]. In particular, LI19 considered 120 simulation runs of truncated disks, each of which started with 20 embryos and 7000 planetesimals. In modelling truncated disks, they tested dependence on core region size (0.7–1.0 vs. 0.7–1.2 au), the distribution of disk mass in between embryos and planetesimals (ratio; $r = 1$, 4, and 8), and the influence of low-mass disk components beyond 1.0 (or 1.2) au. LI19 also proposed a rigorous classification system to properly identify the analogues of Mercury, Venus, Earth, and Mars in a given system (see Section 2 in that study for details). LI19 concluded that truncated disks probably cannot explain the formation of the terrestrial planets for the following main reasons. First, the analogues of Mercury and Mars were more massive, too dynamically cold, and formed too close to the Venus and Earth analogues.

---

[2] In this work, "truncated disks" refers to the disks obtained in Grand Tack models at the end of gas dispersal, whereas "narrow disks" refers to the disk postulated by H09 and its later replications (e.g. Raymond & Izidoro 2017 and Clement et al. 2019a).

[3] In this work, planet analogues are defined as planets with orbits *and* masses similar to those of the terrestrial planets. More details are given in Section 2.

Second, the Earth analogues experienced Moon-forming giant impacts too early and accreted too much mass after those impacts. Nevertheless, LI19 did not explicitly consider narrow disks as modelled in H09. Also, the dependence of their results on the number of embryos used in the disk core region has not been well explored. Thus, it is important to revisit terrestrial planet formation in narrow disks.

The main goal of this study was to investigate whether narrow disks are capable of forming the four terrestrial planets consistently. In particular, we determined the likelihood of such disks producing three- or four-planet analogue systems. We required our planet analogues to belong to those systems and scrutinised their properties against fundamental success criteria, as discussed below. To address these questions, we performed N-body simulations of terrestrial planet formation based on narrow disks. Major improvements over H09 include the following: 1) better number statistics: we simulated the original H09 disk model using 200 runs and five new, similar disk models using 50 runs each; 2) higher disk resolution: 10,000 planetesimals were considered in the new disk models, so dynamical friction was properly modelled; 3) accurate statistical analyses of final systems: our classification algorithm allowed us to properly identify terrestrial planet analogue systems and determine properties of the planets formed in those systems; and 4) the influence of embryo number (mass) and presence of planetesimals in the disk: we varied the number of embryos in the new disk models tested. We also included planetesimals in addition to embryos in some of these models.

## 2. METHODS

We considered a primordial solar system consisting of Jupiter, Saturn, and a protoplanetary disk containing several embryos and residual planetesimals. As in H09 and other similar works (Raymond et al. 2009; Izidoro et al. 2015; Clement et al. 2019a), we did not consider disk gas dynamics in this study. Uranus and Neptune were also not considered because their influence was likely negligible (Walsh et al. 2011; LI19). The initial conditions described here were created following substantially the same methods as found in those works.

Jupiter and Saturn started on their current orbits in all disk models. This choice implicitly assumes that any giant planet instability occurred early in the history of the solar system (within ~10 Myr), so the giant planets did not migrate in our simulations. Indeed, there is growing evidence that the instability occurred early in the solar system's history (Kaib & Chambers 2016; Quarles & Kaib 2019; de Sousa et al. 2020). Because such instability is unimportant for narrow or truncated disks (Deienno et al. 2018; Clement et al. 2019a), our initial conditions are justified, and at the same time, our model remains as simple as originally envisioned by H09. In addition, our choice of the giant planet architecture above is supported by the results of LI19, who found that Jupiter and Saturn in more eccentric orbits tend to produce terrestrial planet systems more similar to our own.

We modelled our narrow disks by initially placing the embryos and planetesimals within 0.7–1.0 au. In particular, we varied the number of embryos from 20 to 10,000 in six disk models and included 10,000 planetesimals in four of those disks (see Table 1 for details).

Irrespective of the number of embryos or planetesimals in the disk, all disks started with a total mass equal to 2 $M_\oplus$. In disks that included planetesimals, embryos and planetesimals carried 80% and 20% of the disk mass, respectively (i.e., $r = 4$). The choice of $r = 4$ was motivated by the findings of recent terrestrial planet studies (e.g. Jacobson & Morbidelli 2014; LI19) and models including disk gas (Morishima et al. 2010; Walsh & Levison 2019; Clement et al. 2020) that favoured protoplanetary disks with $r \geq 4$. In a given disk model, the mass of the individual embryo was the same. We verified the influence of embryo mass by testing distinct numbers of embryos, as described above. Depending on the disk model used, embryos started with masses comparable to the range Ceres–Mars. The initial values of $a$, $e$, and $i$ for the embryos and planetesimals were chosen randomly within the following ranges[4]: embryos: $a_0 = 0.7$–1.0 au, $e_0 < 0.01$, $i_0 \approx e_0/2$ (~0.3°); planetesimals: $a_0 = 0.7$–1.0 au, $e_0 < 0.02$, $i_0 \approx e_0/2$ (~0.6°) (Kokubo & Ida 2000; Morishima et al. 2010). The bulk density of both embryos and planetesimals was initially 3 g cm$^{-3}$.

We performed a total of 450 N-body simulations to investigate the disk models described above. We used a modified version of the MERCURY integrator to execute the simulations (Chambers 1999; Hahn & Malhotra 2005; Kaib & Chambers 2016). Following Brasser et al. (2016a), the integrator calculated the bulk density/radii of forming planets as a function of mass to approximately mimic those of the real planets. The giant planets and embryos gravitationally perturbed one another. The planetesimals interacted with the planets or embryos but did not mutually perturb one another. Collisions were considered 100% mergers, which is an acceptable simplification (see LI19 for a detailed reasoning). All simulations were evolved until 1 Gyr. The time step was 4.57 days (1/80 year), which is small enough to allow reliable calculations for planets with orbits similar to that of Mercury. Bodies that acquired heliocentric distances smaller than 0.15 au or larger than 20 au were eliminated from the simulations to avoid inaccurate orbital evolutions near the Sun and to save time computing distant orbits, respectively. Regardless, these bodies can collide with the Sun or be gravitationally ejected by Jupiter on very short timescales, so their influence on our results is negligible. A summary of the initial conditions is given in Table 1.

In this work, we used our classification algorithm to identify all planet analogues formed in a given system. First the algorithm identified the Venus and Earth analogues of the system (typically the two most massive planets formed within 1.5 au). Then the region inside the Venus analogue and the one outside the Earth analogue defined the Mercury and Mars regions, respectively. If the Venus–Earth pair was not found in the system, the classification stopped, and the system was deemed unsuccessful for the purposes of this work. Next the algorithm identified the Mercury and Mars analogues (if they existed) inside these regions. The standard planetary mass ranges considered for analogue candidates were 0.03–0.17 (Mercury), 0.4–2.0 (Venus and Earth), and 0.05–0.32 $M_\oplus$ (Mars), but we also tested more stringent criteria defined by 0.03–0.11, 0.6–1.5, and 0.05–0.21 $M_\oplus$, respectively. A planet was defined as any object with mass m $\geq$ 0.03 $M_\oplus$ (a minimum of 50% of the mass of Mercury). The details of our classification algorithm and the importance of properly identifying planet analogues in a given system are discussed in LI19.

---
[4] The 20 embryos of disk model D1 (Table 1) were distributed uniformly at 0.7–1.0 au to avoid potential large variation in initial mutual perturbations among the embryos at the start of the simulations.

The results and discussion in this work are based solely on three- or four-planet analogue systems and the planets formed in those systems as identified by our classification algorithm. This includes the evaluation of model success based on five main constraints (for detailed discussions, see Clement et al. 2018; LI19; Raymond et al. 2020).

**A**. **Formation of Mercury, Venus, Earth, and Mars analogues**. Because it is difficult to produce the four terrestrial planets simultaneously (LI19), we required a minimum of three planet analogues to form in a given final system. An analogue system had to contain Venus, Earth, and at least a third analogue (Mercury or Mars) in the same system. Each planet also had to have $a \leq 2$ au.

**B**. **Terrestrial planet system properties**. The angular momentum deficit (AMD) measures the dynamical excitation of a system, and planets on more eccentric or inclined orbits increase the AMD of a system. The radial mass concentration (RMC) evaluates the mass distribution of the planets in a system, so it can describe how the system total mass is concentrated in terms of orbital range and number of planets. Finally, the orbital spacing of planets (OS) reveals the degree of spatial compactness of the planets formed in a system (see Chambers 2001 for more details). The final systems should possess 50–200% of the RMC and OS and a maximum of 200% of the AMD of the solar system's terrestrial planets (Jacobson & Morbidelli 2014).

**C**. **Timing of the giant impact that formed the Moon**. Earth probably experienced this giant impact within time $t \approx 20$–140 Myr in the framework of our simulations (Albarede 2009; Brasser et al. 2016a and references therein). Here, the collision of an impactor with at least 10% of the target's mass was considered a giant impact (Canup 2012; Jackson et al. 2018).

**D**. **Late veneer mass delivered to Earth**. The mass delivered to Earth via impacts of remnant embryos or planetesimals after the planet's last giant impact probably comprises 0.1–1% of Earth's mass (Raymond et al. 2014; Brasser et al. 2016b).

**E**. **Formation timescales of the four terrestrial planets**. The Venus–Earth pair likely assembled on a 100-Myr timescale (Morbidelli et al. 2012; Raymond et al. 2014), whereas Mars probably acquired its bulk mass within $t \approx 10$ Myr (Dauphas & Pourmand 2011; Jacobson & Morbidelli 2014; McCubbin & Barnes 2019). The formation timescale of Mercury is unknown.

## 3. MAIN RESULTS AND DISCUSSION

### 3.1 Terrestrial planet analogue systems

We considered all planets obtained in three- and four-planet analogue systems when computing system properties. Table 2 presents the median key properties of the disks modelled in this work. Note that we tested two classification criteria to identify our planet analogues. However, we found that the results were quite similar for both criteria, and thus we opted for the one that yielded larger statistics (standard criteria). For completeness, we have summarised the results using the stringent criteria in supplementary tables in the Appendix.

We identified 72, 24, 19, 19, 18, and 12 analogue systems (164 in total) in the D0–D5 disk models tested, respectively. If we consider the number of initial simulation runs per disk, we obtained analogue system fractions of 36, 48, 38, 38, 36, and 24%, respectively, for the same disks. Given the similarity of these fractions, apparently the initial disk conditions played a minor role in forming the system analogues. Also, as required by the classification algorithm, the Venus–Earth analogue pair was present by default in all of those systems. Most of the obtained analogue systems consisted of Venus, Earth, and Mars analogues. By contrast, only eight analogue systems contained the four terrestrial planets in the same system (a 4-P system) (Figure 1). Thus, our narrow disks produced 4-P systems with efficiencies $8/164 \approx 5\%$ per analogue system or $8/450 \approx 2\%$ in general (i.e., the success rate for constraint A). In addition, although all disk models formed a significant number of Mars analogues, there were only 24 identified Mercury analogues for all disks combined ($24/450 \approx 5\%$ production efficiency). In agreement with the findings of LI19, these results suggest that it is difficult to form both Mercury analogues and 4-P systems.

In general, the system analogue fractions described above were comparable to those found in LI19 for truncated disks with a disk mass concentrated at 0.7–1.0 au (43%; *T7-10* disks in that work). Also, Clement et al. (2019a) found 58% and 22–29% analogue system fractions for H09's original narrow disk and narrow disks perturbed by giant planet instability, respectively[5]. So, despite the model or classification differences between the work of Clement et al. (2019a) and LI19 and the different initial conditions in narrow and truncated disks as considered in LI19 and this work, we found similar analogue system fractions. This suggests that fundamental properties of the disk, such as the size of the core region where most of the disk mass is concentrated, whether the disk mass is dominated by embryos or planetesimals, and the orbital architecture of the giant planets, determine the final outcomes of analogue systems.

Regarding system properties (constraint B), the analogue systems obtained in D0 disks were overly excited, which suggests that the dynamical friction in such disks was not strong enough to counterbalance the self-stirring caused by the mutual gravitational interactions of 400 embryos. The inclusion of planetesimals increased the effects of dynamical friction, thus reducing the system excitation to values comparable to or about twice those of the solar system (Table 2). Note that the ratio of individual embryo-to-planetesimal mass was large enough for dynamical friction to operate in all disks. Nonetheless, systems dynamically comparable to those of the solar system were commonly obtained for disk D1 (median AMD ~1.5 times that of the solar system), which was the disk with the smallest initial number of embryos. In particular, in those disks there was less gravitational mutual excitation or scattering among the embryos from the start, and the ratio of individual embryo-to-planetesimal mass was the largest among our disks. Thus, those features likely enhanced dynamical friction, resulting in systems with planets with colder orbits at the end of the simulations. In fact, only systems formed in disk D1 produced a significant number of Venus and Earth analogues with cold orbits comparable to those of the real planets (Table 3).

---

[5] Note that these results were obtained for systems that evolved until 200 Myr. Some of those systems may still be evolving dynamically after several hundred Myr.

In addition, although the orbital compactness of the final systems was remarkably similar to that of the solar system in all disk models (OS in Table 2), the orbits of the planet analogues were somewhat different. That is, in general Mercury analogues formed closer to Venus analogues than in reality, and Venus and Earth analogues typically formed not as close to each other as observed (Table 3). These orbital trends and the fact that our Earth analogues were systematically less massive than Earth also resulted in RMC values slightly smaller than those of the solar system. Last, the distribution of planet analogues in orbital-mass space surprisingly reveals the inability of our narrow disks to reproduce both the orbits and masses of the Venus–Earth pair (Figure 2). This failure is clearly related to the disk initial conditions. Although recent terrestrial planet formation models advocate the need for similar narrow/truncated disks to explain Mars' small mass, the aforementioned results suggest the need to revisit the initial conditions used in this kind of disk.

Regarding the last giant impact timing and late veneer mass fraction of the Earth analogues, we found that in general the former occurred before the minimum 20 Myr (constraint C) for 60–75% of Earth analogues identified in all disk models (in agreement with Clement et al. 2019a). Also, as a consequence of this relatively fast accretion of Earth analogues, most of them accreted too much late veneer mass. Overall, only 0–15% and 30% of these analogues satisfied the 0.1–1% constraint (D) for disks D0/D2–D5 and D1, respectively. These results are summarised in Figure 3. In particular, disk D1 produced systems in which the Earth analogues had the potential to satisfy both constraints C and D. In particular, 8 of 24 Earth analogues achieved this. Disk D1 started with embryos nearly as massive as Mars, so they quickly and strongly perturbed the disk planetesimals in the simulations. The objects in the disk were also excited by the resonant and secular perturbations of Jupiter and Saturn because both planets started on their current (slightly eccentric) orbits in our simulations (Haghighipour & Winter 2016; LI19 for more details). For these reasons, fewer remnant planetesimals could be accreted as late veneer mass to the newly formed Earth analogues, hence the higher success in satisfying constraint D. Similarly, in such hot disk conditions the accreting collisions took longer on average to occur because of their lower probabilities. As a result, Earth analogues finished forming later. Note that the success rates of constraints C and D in disk D1 (40% and 30%, respectively) were significantly higher than those found in LI19's *T7-10* disks (<10% and <5%, respectively). *T7-10* disks were modelled with Jupiter and Saturn placed on near-circular orbits at 5.5 and 7.5 au, thus with much less perturbative influence on the disk. Besides, the outer region of *T7-10* disks also contributed to late mass accretions of materials to the Earth analogues, thus worsening the success rate of constraint D. In sum, a dynamically excited disk tends to lengthen Earth's formation and decrease its mass accreted by remnant planetesimals.

However, the success rates concerning constraint C should be considered with caution. First, the minimum Moon-forming giant impact of 20 Myr is based on the lower end value of 30 Myr as found by hafnium-tungsten chronology (Kleine et al. 2005; Albarede 2009) assuming that the solar system nebular gas dissipated in 10 Myr (our time zero). Had it dissipated earlier, in 3–5 Myr, as found in the literature (e.g. Walsh et al. 2011; Chambers 2016; Wang et al. 2017), the giant impact timing would increase to 25–27 Myr, thus reducing the success rates discussed above. Second, more recent studies based on the same chronology suggest that the Moon formed 50 Myr after the solar system (Thiemens et al. 2019), or approximately 40–47 Myr after gas dissipation. For instance, a minimum timing of 40–47 Myr would reduce the number of successful Earth analogues in disk D1 to 4–6 out of 24, so the success rate would decrease to 15–25%.

Regarding the Mars formation timescale, we found that disk D1 was very successful in producing systems in which the majority of Mars analogues were able to form within the maximum ~10 Myr (90% of these planets satisfied constraint E). The remaining disks, D0, D2, D3/D4, and D5, yielded approximate success rates of 60, 70, 50, and 40%, respectively. As expected, these differences reflect the distinct initial embryo number, and therefore mass, considered in each disk model. In particular, whereas the Mars analogues in disk D1 started as embryos with 47% of their final masses[6], their counterparts in disks D2, D0/D3, and D4/D5 started with 14, ~4–5, and ~0.1–0.5%, respectively. Thus, Mars analogues in disk D1 managed to acquire masses comparable to that of Mars in a shorter amount of time and with fewer giant impacts (Table 3).

In summary, if the early inner solar system is represented by a narrow disk, these combined results suggest that the disk contained a few tens of embryos with masses comparable to that of Mars embedded in a low-mass sea of planetesimals and the giant planets in their current orbital configuration. Because embryo/planetesimal formation models can produce such conditions in a few Myr (Chambers 2016; Walsh & Levison 2019), our results also suggest that the giant planets acquired their current orbits on similar time scales, which supports the preferred instability timing of ~1–5 Myr in Clement et al. (2018, 2019a).

### 3.2 Planet analogues

We present below the main properties of our analogues of Mercury, Venus, Earth, and Mars formed in three- or four-terrestrial-planet analogue systems. The details are summarised in Table 3 and Figure 2. In the discussions below, we emphasise the results and implications for disk D1, because it yielded systems containing planet analogues with better prospects for reproducing the real planets.

*Mercury*. We obtained 10, 1, 3, 1, 6, and 3 Mercury analogues in disks D0, D1, D2, D3, D4, and D5, respectively. If we consider the identified analogue systems for each model, the production fraction of Mercury analogues per analogue system becomes approximately 15, 5, 15, 5, 33, and 25%, respectively. Thus, a greater initial number of small embryos might increase the odds of obtaining Mercury analogues in such systems. However, recall that these production fractions refer to analogue systems, so they drop to 5% if we consider narrow disks in general or 2% if we generalise the results solely to 4-P systems (see Section 3.1).

---

[6] Indeed, the contribution from embryos and planetesimals to the remaining final mass of Mars analogues in disk D1 was approximately 45% and 8%, respectively.

Reproducing Mercury's excited orbit at $a \sim 0.39$ au and its low mass at the same time is perhaps the most difficult constraint to satisfy among the four terrestrial planets[7] (Lykawka & Ito 2017). Here our Mercury analogues may be considered comparable to Mercury in terms of mass, but their orbits were in general slightly less excited than that of the real planet. However, more important, these analogues tended to form farther from the Sun (at median $a \approx 0.46$ au) *and* closer to Venus analogues than in reality (~0.15–0.21 au vs. 0.34 au)[8]. Thus, although we note an improvement in terms of orbital excitation of our Mercury analogues compared to LI19's results in *T7-10* disks, our narrow disks have serious difficulty reproducing Mercury (Figures 2 and 4).

Mercury's large iron core may be better explained if the planet suffered multiple giant impacts during its formation (Asphaug & Reufer 2014; Jackson et al. 2018). Overall, our Mercury analogues experienced two or three giant impacts during their formation, which may support that scenario. However, note that N-body simulations of narrow disks that included fragmentation and considered the internal structure of planets failed to reproduce the large iron core and the current orbit of Mercury (Clement et al. 2019b).

*Venus*. We obtained 164 Venus analogues from the results of disks D0–D5 combined. The cold orbit of Venus was reproduced statistically in disk D1 but only marginally in the other remaining disks. As discussed in the previous section, this highlights the importance of dynamical friction in forming a planet with low $e$ and $i$ (see also O'Brien et al. 2006). Although our Venus analogues were in general as massive as the real planet, they tended to form significantly closer to the Sun (at median $a \approx 0.61$ au) *and* farther from Earth analogues than in reality (0.34 au vs. 0.28 au) (Figure 2).

Venus analogues experienced five to nine giant impacts during their formation histories when we consider all disks combined. This may support the hypothesis of the absence of Venusian satellites via fortuitous giant impacts (Alemi & Stevenson 2006). However, the lack of a magnetic dynamo in Venus's core may be difficult to explain if it is required that the planet not have experienced such giant impacts during its formation (Jacobson et al. 2017).

*Earth*. As our classification algorithm required the presence of the Venus–Earth analogue pair in the final system, we also obtained 164 Earth analogues from all disks combined. In agreement with LI19's results, we found that the formation and several key properties of our Earth analogues were similar to those of their Venus counterparts. For instance, Earth analogues with $e$ and $i$ comparable to those of our planet were more commonly obtained in disk D1. The masses of Venus and Earth analogues were also comparable. However, in general our

---

[7] Roig et al. (2016) found that Mercury's eccentricity and inclination could be produced in the framework of the "jumping-Jupiter" instability model. However, they assumed that Mercury was already formed at $a = 0.387$ au with its current small mass. In addition, their obtained Venus- and Earth-like planets are too excited dynamically.

[8] We obtained 0.15 and 0.21 au distances by subtracting the planet analogues' $a$ for all disks combined (in Table 3) and calculating the median of $a_{Venus} - a_{Mercury}$ over individual systems that contained both planets. Based on the same data, the ratios of orbital periods for Venus and Mercury analogues are $(0.61/0.46)^{1.5} = 1.53$ and 1.75 (median of $(a_{Venus}/a_{Mercury})^{1.5}$ over the individual systems). These ratios are notably smaller than the observed $(0.723/0.387)^{1.5} \sim 2.55$.

Earth analogues were systematically less massive than Earth (Figure 2). When we analysed each disk individually, the variation in mass for the Earth analogues was approximately ±20%–30% about the median; thus, our narrow disks reproduced Earth's 1 $M_\oplus$ only marginally. Nevertheless, 10 of the 24 Earth analogues obtained in disk D1 acquired masses within the 0.8–1.2 $M_\oplus$ range. Finally, overall our Earth analogues often formed near Earth's current location at $a = 1$ au with 1-sigma variation less than ±0.1 au in each disk model (at median $a \approx 0.95$ au).

Similar to the results found for Venus analogues, our Earth analogues experienced five to nine giant impacts during their formation. However, the timing of the last giant impact experienced by these analogues was earlier than the less conservative minimum timing of 20 Myr for 60–75% of Earth analogues. For a more realistic timing of 40 Myr, the fraction would reach 75–95%. In particular, disk D1 was responsible for the lower end values of these fractions. Also, 4 of the 10 Earth analogues that acquired 0.8–1.2 $M_\oplus$ masses in disk D1 experienced their last giant impacts 37, 41, 60, and 80 Myr after the start of the simulations. Therefore, our results indicate that such late giant impacts would occur for roughly 15% of our best Earth analogues.

Regarding the delivery of water to Earth by volatile-rich objects in the protoplanetary disk, based on truncated disks under the Grand Tack scenario, O'Brien et al. (2014) found that enough water could be delivered to Earth if volatile-rich planetesimals were present beyond the disk core region at 0.7–1.0 au. As typically found in Grand Tack simulations, remnant planetesimals are found at ~1.0–3.5 au, and a substantial fraction of them are akin to C-type asteroids that possess high water mass fractions (referred to as "C-planetesimals" below). Using similar truncated disks, LI19 confirmed that the presence of C-planetesimals could explain Earth's water budget. However, they also found that the same C-planetesimals would deliver too much water to Venus and Mars, by a factor of at least a few times the upper limits of estimated water budgets in those planets during the early solar system. In addition, from the comparison of *T7-10* and narrow disks (Section 3.1), we found that the presence of C-planetesimals can increase late accreted mass to Earth, thus significantly decreasing the success rate of the late veneer mass constraint. Here this constraint may be marginally satisfied in disk D1, but Earth's water constraint would not, because C-planetesimals are not considered in narrow disks (only in truncated disks). In conclusion, it is unlikely that both the late veneer mass fraction and the Earth's water budget constraints can be simultaneously satisfied by Earth analogues produced in either truncated or narrow disks.

*Mars*. As a result of the absence of mass beyond 1 au, it is not surprising that our narrow disks produced a large number of Mars analogues (148 in total). The production fraction of Mars analogues per analogue system was approximately 75–95%, or ~33% (148/450) for narrow disks in general. Thus, in line with previous work, relatively small-mass Mars analogues are often obtained in narrow disks.

Can narrow disks reproduce Mars's moderately excited orbit at $a \sim 1.52$ au and its low mass at the same time? The answer might be yes. Although our Mars analogues formed closer to the Sun than in reality (at median $a \approx 1.46$ au), they acquired orbits comparable to that of Mars irrespective of the disk model, particularly in terms of *e* and *i*. Furthermore, although these analogues were somewhat more massive than Mars, overall their individual disk median masses were within a factor of ~2 that of Mars. In short, consistent with published Mars formation models (footnote 1), our narrow disks have the potential to reproduce Mars's orbit and small mass at the same time (Figure 2).

The Mars analogues obtained in this work possessed smaller masses and orbits more similar to those of Mars than the analogues found in *T7-10* truncated disks (LI19). Because the embryos and planetesimals had similar initial conditions for narrow and truncated disks, we believe that the main reason for this improvement was that Jupiter and Saturn were considered on their current orbits in narrow disks, rather than on low-$e$ or -$i$ pre-migration orbits, as typically modelled in truncated disks (Walsh et al. 2011; Jacobson & Morbidelli 2014; Walsh & Levison 2016). Because both giant planets have higher eccentricity in narrow disks, the associated planetary secular or resonant perturbations were much stronger in the outer regions of the disks. Therefore, these effects likely inhibited the formation of more massive Mars analogues and reduced the effects of dynamical friction in the disk beyond ~1 au, thus allowing these planets to acquire more excited orbits.

Last, our Mars analogues typically experienced one to four giant impacts by the end of their formation. This may support global events such as the one that caused the surface dichotomy on Mars (Marinova et al. 2008).

## 4. SUMMARY

We performed N-body simulations of terrestrial planet formation to test whether narrow protoplanetary disks were capable of producing systems and planets analogous to our own. Out of 450 simulation runs, we found a total of 164 terrestrial planet analogue systems, of which 24 and 148 contained analogues of Mercury and Mars, respectively. Only eight analogue systems were 4-P ones.

We scrutinised the properties of the planet analogues identified in our analogue systems against five main constraints in the inner solar system: *A*. formation of Mercury, Venus, Earth, and Mars analogues, in particular their orbits ($a$, $e$, $i$), proximity to Venus or Earth, and masses for the Mercury and Mars analogues; *B*. terrestrial planet system properties, in particular the system dynamical state as measured by AMD; *C*. timing of the Moon-forming giant impact within 20–140 Myr; *D*. late veneer mass delivered to Earth within 0.1%–1%; and *E*. formation timescales of the four terrestrial planets, in particular that of Mars within ~10 Myr. We also compared our analogue systems with those found for truncated protoplanetary disks, which considered distinct giant planet orbital configurations and low-mass components beyond ~1 au in previous works.

The main findings of this work indicate that:
1. The probabilities of producing Mercury and Mars analogues in narrow disks are 5 and 33%, respectively. However, the production rates drop to 2% if we consider solely 4-P systems.
2. Narrow disks are incapable of reproducing the orbits and masses of the four terrestrial planets consistently. In particular, the following serious shortcomings stand out:
- The formation of Mercury analogues was too inefficient, and even those that formed were

located farther from the Sun and too close to Venus analogues compared to the real planets.

- Venus and Earth analogues failed to reproduce the Venus–Earth pair in orbital-mass space. In particular, the mutual orbital spacing of Venus and Earth analogues was systematically larger than in reality. Earth analogues were also systematically less massive than our planet (median 0.79 $M_\oplus$).
- Venus-to-Earth mass ratios were incorrect in ~75% of the analogue systems (outside the interval 0.6-1.0).

3. Narrow disks produce analogue systems that are more excited than the inner solar system. However, disks that started with 20 massive embryos (totalling 1.6 $M_\oplus$) embedded in a sea of small-mass planetesimals yielded better prospects for reproducing the dynamical state of our system. This highlights the importance of dynamical friction during terrestrial planet formation.

4. Initial embryos with individual masses comparable to that of Mars are necessary to provide adequate matches to Earth's accretion history. Specifically, massive initial embryos perturbed nearby planetesimals on to orbits that are more dynamically excited, thereby lengthening Earth's accretion timescale and limiting the amount of planetesimal mass available to form the late veneer.

5. To form Mars on timescales consistent with geochemical constraints, more massive initial embryos are preferred in narrow disks (>0.02 $M_\oplus$).

6. Jupiter and Saturn probably had already acquired orbits comparable to their current ones by the onset of terrestrial planet formation. This orbital configuration allowed mean motion and secular resonances associated with the giant planets to excite the orbits of embryos/planetesimals in the disks. As a result, this perturbation not only enhanced the mechanisms described in item 4 but also increased the odds of reproducing Mars' excited orbit.

In conclusion, narrow disks evidently face major outstanding problems that still must be resolved to explain the formation of the four terrestrial planets. Nevertheless, these disks could be considered the baseline for improved disk models that aim to satisfy the main constraints in the inner solar system. For instance, as discussed in LI19, our results strengthen the case that the terrestrial planets and the asteroid belt may have formed from a disk that contained mass concentrated within a core region (e.g. 0.7–1.0 au) surrounded by mass-depleted inner and outer region components.

## ACKNOWLEDGMENTS

I would like to thank the reviewer Matthew S. Clement for a number of insightful comments, which improved the overall presentation and flow of this work. I appreciate the help of Takashi Ito with gnuplot in preparing this work. All simulations presented here were performed using the general-purpose PC cluster at the Center for Computational Astrophysics (CfCA) in the National Astronomical Observatory of Japan (NAOJ). I am thankful for the generous time allocated to run the simulations.

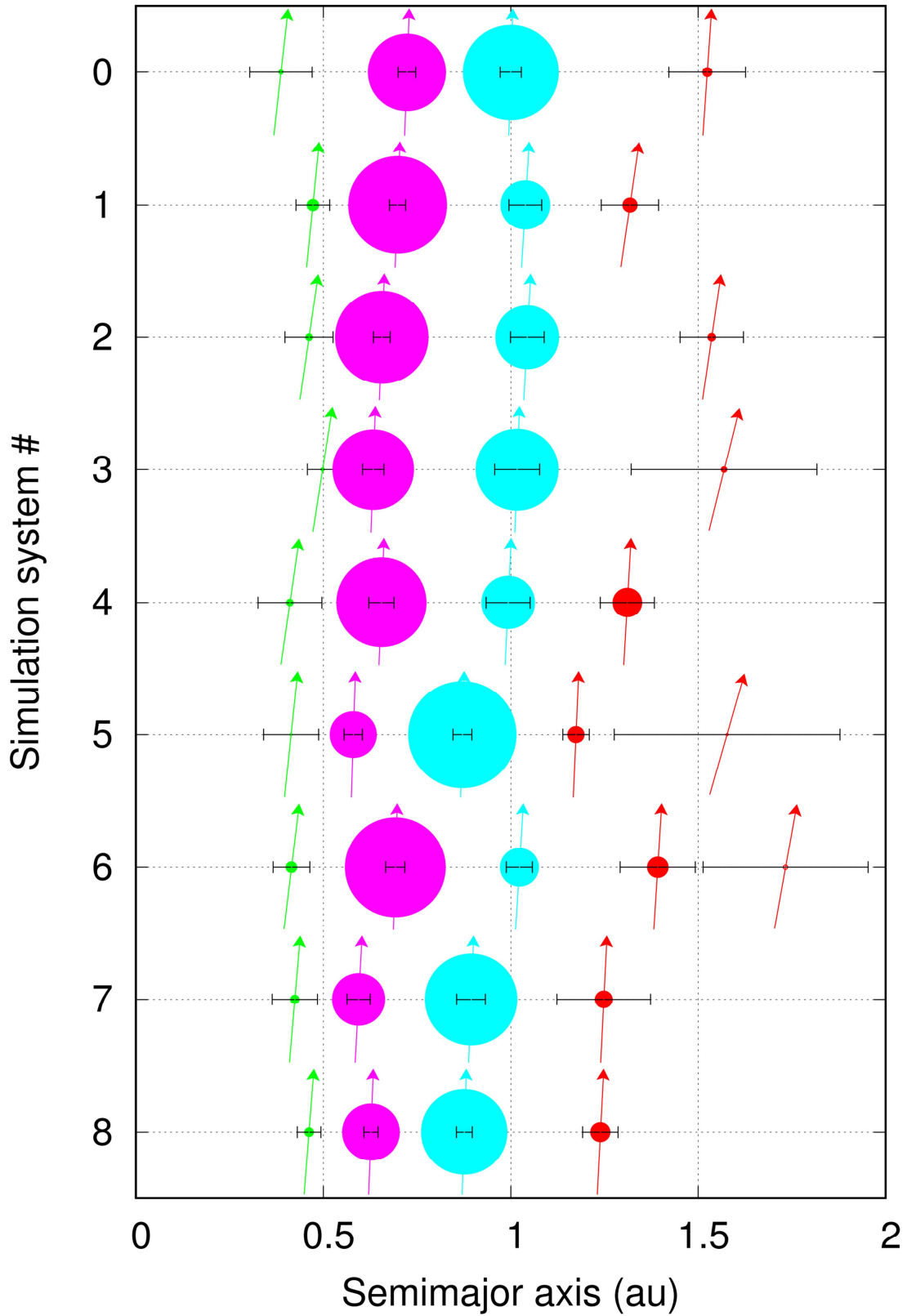

**Figure 1.** A comparison of individual 4-P analogue systems (#1–8) with the planets of the solar system ("system #0" shown at the top). The data are considered according to D0 (#1–4), D2 (#5–6), and D4 (#7–8) disks. Note that systems #5 and #6 contain additional Mars-like planets. The inclination $i$ of the planets is represented by the angle between the vector and the perpendicular (e.g. the vector points to the top for $i = 0°$). The variation in heliocentric distance based on the object's perihelion and aphelion is indicated by the error bars. The size of a planet scales proportional to its mass.

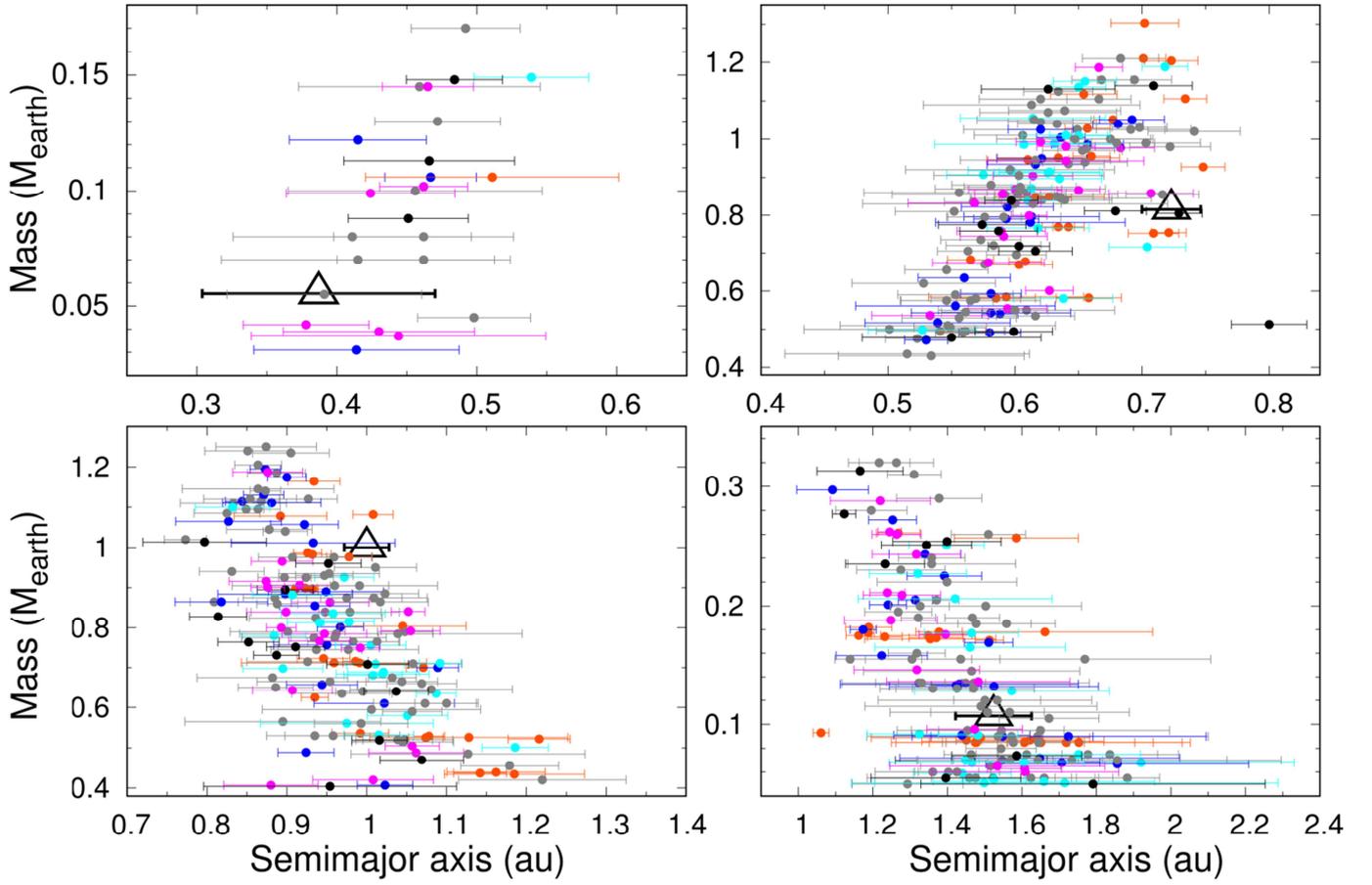

**Figure 2**. Representative analogues formed in three- or four-terrestrial-planet analogue systems obtained from all simulations: 24 Mercury analogues (top left), 164 Venus analogues (top right), 164 Earth analogues (bottom left), and 148 Mars analogues (bottom right). Further details are given in Section 3.2 and Table 3. The variation in heliocentric distance based on the object's perihelion and aphelion is indicated by the error bars. The different colours indicate the disk model used: D0 (grey), D1 (orange), D2 (blue), D3 (cyan), D4 (magenta), and D5 (black). The large open triangles represent the terrestrial planets of the solar system.

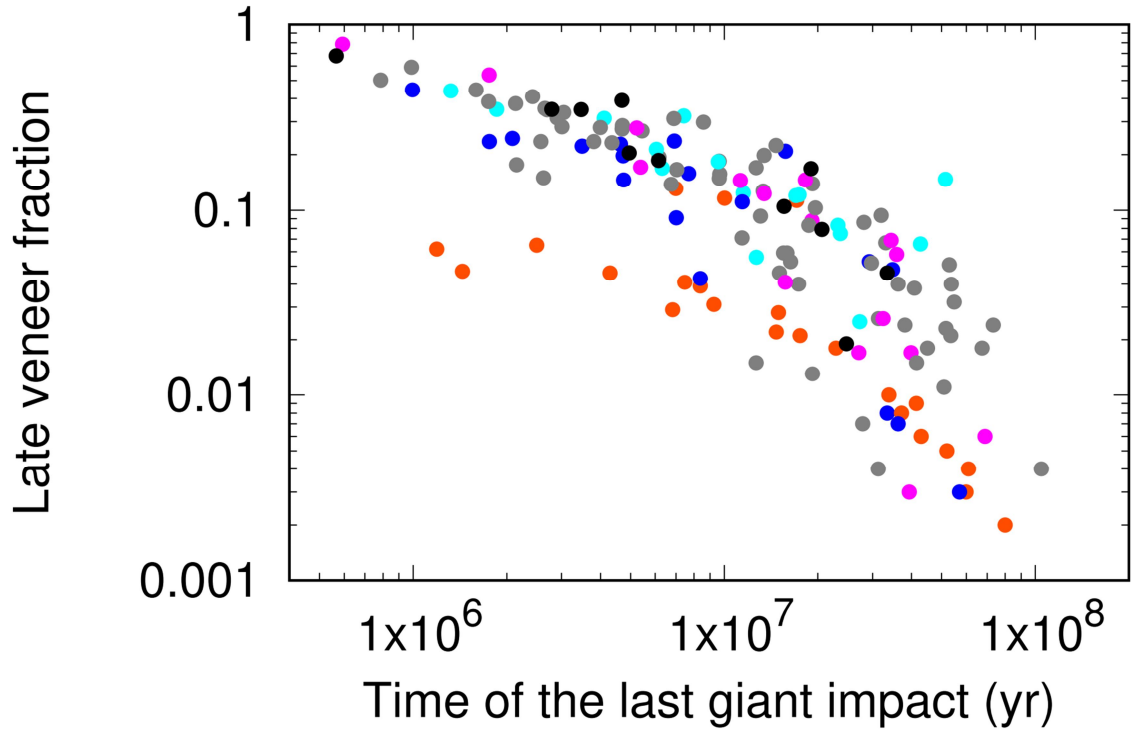

**Figure 3.** Time of the last giant impact experienced by 149 Earth analogues that accreted at least 0.1% mass thereafter (minimum identifiable fraction). These analogues formed in three- or four-terrestrial-planet analogue systems obtained from all simulations. The different colours represent the disk models tested in this work, as explained in the caption of Figure 2. See Section 3.1 for more details.

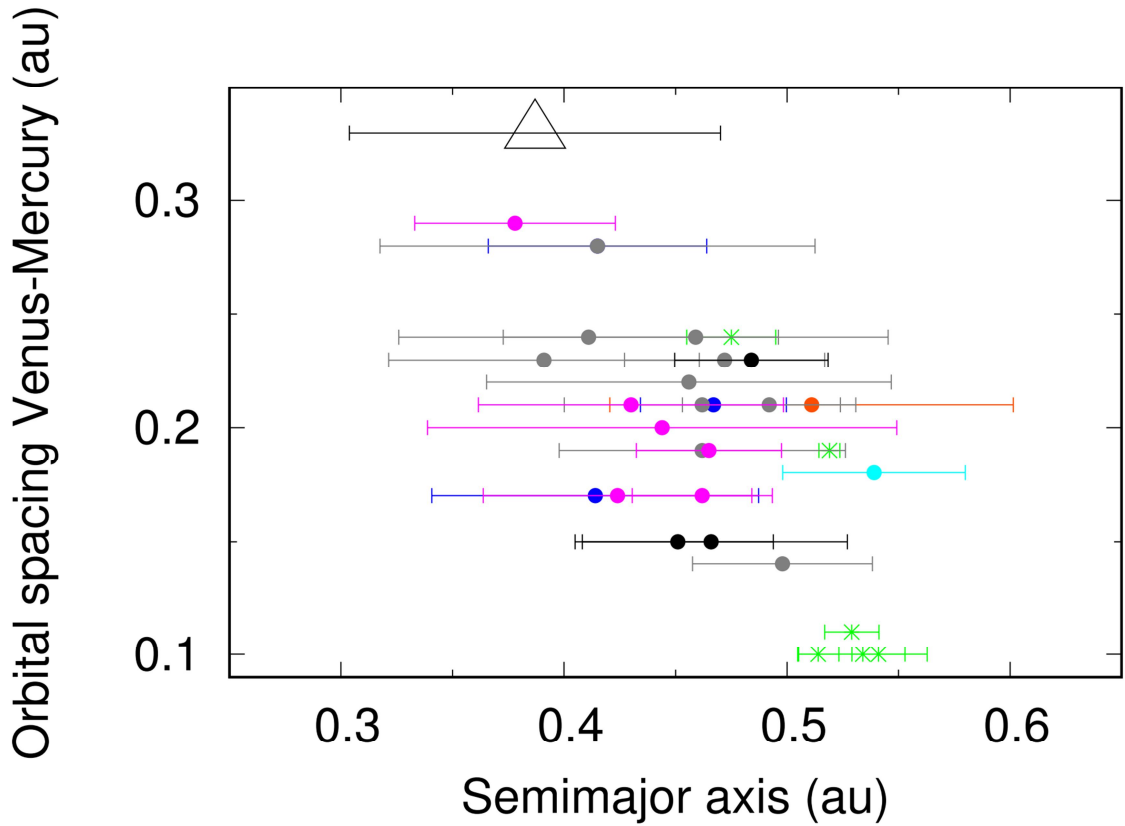

**Figure 4.** Our 24 representative Mercury analogues formed both farther from the Sun and closer to the Venus analogues in their systems compared to reality. These analogues formed in three- or four-terrestrial-planet analogue systems obtained from all simulations. Six Mercury analogues obtained in Lykawka & Ito (2019) for *T7-T*0 disks are also shown with green asterisks. The other symbols/colours have the same meaning as those in Figure 2. See Section 3.2 for more details.

**Table 1.** Initial conditions for the narrow protoplanetary disks in our simulations

| Disk | Number of embryos [Mass of each embryo ($M_\oplus$)] | Planetesimals | $r$ | # runs |
|---|---|---|---|---|
| D0 | 400 [0.005] | 0 | … | 200 |
| D1 | 20 [0.08] | 10,000 | 4 | 50 |
| D2 | 80 [0.02] | 10,000 | 4 | 50 |
| D3 | 400 [0.004] | 10,000 | 4 | 50 |
| D4 | 2000 [0.0008] | 10,000 | 4 | 50 |
| D5 | 10,000 [0.0002] | 0 | … | 50 |

**Notes.** $r$ represents the ratio of total disk mass in embryos to that in planetesimals.

All disks started with a total disk mass of 2 $M_\oplus$ distributed in embryos and planetesimals confined at 0.7–1.0 au.

The giant planets Jupiter and Saturn started on their current orbits.

Individual masses of embryos in disks D1, D0/D2/D3, and D4/D5 are comparable to the masses of Mars, the Moon, and Ceres, respectively.

Individual masses of planetesimals are 0.00004 $M_\oplus$ (~1/4 Ceres).

See Section 2 for more details on the disk models.

**Table 2.** Properties of the three- or four-terrestrial-planet analogue systems

| Disk Model (# runs) | n3$_{mve}$ | n3$_{vem}$ | n4 | AMD | RMC | OS | Lvf (%) | MVr | VEr | MEr |
|---|---|---|---|---|---|---|---|---|---|---|
| D0 (200) | 6 | 62 | 4 | 0.0047 | 70.9 | 37.2 | 9.0 | 0.09 | 1.03 | 0.14 |
| D1 (50) | 1 | 23 | 0 | 0.0023 | 74.7 | 32.5 | 2.5 | 0.13 | 1.19 | 0.24 |
| D2 (50) | 1 | 16 | 2 | 0.0036 | 72.5 | 37.9 | 14.6 | 0.13 | 0.89 | 0.17 |
| D3 (50) | 1 | 18 | 0 | 0.0034 | 76.5 | 37.3 | 12.1 | 0.16 | 1.29 | 0.10 |
| D4 (50) | 4 | 12 | 2 | 0.0034 | 74.4 | 34.1 | 6.4 | 0.08 | 1.08 | 0.23 |
| D5 (50) | 3 | 9 | 0 | 0.0027 | 76.5 | 32.8 | 17.7 | 0.15 | 1.03 | 0.32 |
| *T7-10* (60) | 2 | 20 | 4 | 0.0002 | 70.1 | 27.1 | 8.7 | 0.17 | 1.00 | 0.26 |
| **Solar system** | … | … | … | 0.0018 | 89.7 | 37.7 | 0.1–1 | 0.067 | 0.815 | 0.107 |

**Notes.** The number of analogue systems containing Mercury–Venus–Earth, Venus–Earth–Mars, and Mercury–Venus–Earth–Mars representative analogues are given by n3$_{mve}$, n3$_{vem}$, and n4, respectively. AMD is the angular momentum deficit, RMC is the radial mass concentration, OS is the orbital spacing, Lvf is the late veneer mass fraction of the representative Earth analogues, MVr is the ratio of Mercury analogue median mass to Venus analogue median mass, VEr is the ratio of Venus analogue median mass to Earth analogue median mass, and MEr is the ratio of Mars analogue median mass to Earth analogue median mass. Except for the number of analogue systems, all other quantities are represented by medians found in each disk model. The standard planetary mass ranges considered to identify planets in analogue systems were 0.03–0.17 (Mercury), 0.4–2.0 (Venus and Earth), and 0.05–0.32 (Mars) M$_\oplus$. The truncated disk *T7-10* is based on the results of Lykawka & Ito (2019).

**Table 3.** Representative analogues of Mercury, Venus, Earth, and Mars formed in the three- or four-terrestrial-planet analogue systems

|  | N | $a$ (au) | $e$ | $i$ (°) | $m$ (M$_\oplus$) | nGI | tGI (Myr) |
|---|---|---|---|---|---|---|---|
| *Mercury* | | | | | | | |
| D0 | 10 | 0.46 | 0.159 | 8.0 | 0.08 | 2.5 | 28 |
| D1 | 1 | 0.51 | 0.177 | 2.7 | 0.11 | 0 | … |
| D2 | 3 | 0.42 | 0.118 | 7.2 | 0.11 | 2 | 44 |
| D3 | 1 | 0.54 | 0.076 | 8.7 | 0.15 | 3 | 11 |
| D4 | 6 | 0.44 | 0.131 | 5.7 | 0.07 | 1 | 34 |
| D5 | 3 | 0.47 | 0.095 | 7.2 | 0.11 | 2 | 25 |
| All disks | 24 | 0.46 | 0.133 | 7.1 | 0.10 | 2 | 30 |
| | | | | | | | |
| *T7-10* | 6 | 0.52 | 0.023 | 1.5 | 0.13 | 0 | … |
| | | | | | | | |
| *Venus* | | | | | | | |
| D0 | 72 | 0.61 | 0.056 | 3.0 | 0.86 | 9 | 10 |
| D1 | 24 | 0.65 | 0.031 | 1.9 | 0.85 | 6 | 8 |
| D2 | 19 | 0.59 | 0.043 | 2.5 | 0.79 | 8 | 14 |
| D3 | 19 | 0.63 | 0.052 | 2.6 | 0.91 | 8 | 12 |
| D4 | 18 | 0.61 | 0.049 | 2.3 | 0.86 | 6 | 7 |
| D5 | 12 | 0.61 | 0.045 | 2.4 | 0.77 | 5 | 4 |
| All disks | 164 | 0.61 | 0.050 | 2.5 | 0.85 | 7 | 9 |
| | | | | | | | |
| *T7-10* | 26 | 0.63 | 0.007 | 0.5 | 0.76 | 5 | 6 |
| | | | | | | | |
| *Earth* | | | | | | | |
| D0 | 72 | 0.95 | 0.055 | 2.9 | 0.83 | 9 | 15 |
| D1 | 24 | 0.99 | 0.029 | 1.9 | 0.71 | 5 | 16 |
| D2 | 19 | 0.92 | 0.047 | 2.6 | 0.88 | 9 | 8 |
| D3 | 19 | 1.01 | 0.044 | 3.0 | 0.71 | 8 | 17 |
| D4 | 18 | 0.93 | 0.049 | 2.4 | 0.80 | 6 | 23 |
| D5 | 12 | 0.93 | 0.044 | 2.3 | 0.74 | 6.5 | 11 |
| All disks | 164 | 0.95 | 0.046 | 2.6 | 0.79 | 8 | 16 |
| | | | | | | | |
| *T7-10* | 26 | 0.95 | 0.006 | 0.5 | 0.75 | 5 | 5 |
| | | | | | | | |
| *Mars* | | | | | | | |
| D0 | 66 | 1.47 | 0.088 | 8.7 | 0.12 | 3.5 | 19 |
| D1 | 23 | 1.48 | 0.080 | 4.9 | 0.17 | 1 | 4 |
| D2 | 18 | 1.43 | 0.080 | 5.9 | 0.15 | 3 | 44 |
| D3 | 18 | 1.47 | 0.103 | 8.3 | 0.07 | 2 | 9 |
| D4 | 14 | 1.32 | 0.089 | 4.4 | 0.18 | 4 | 15 |
| D5 | 9 | 1.40 | 0.104 | 3.0 | 0.24 | 4 | 8 |
| All disks | 148 | 1.46 | 0.092 | 7.0 | 0.13 | 3 | 15 |
| | | | | | | | |
| *T7-10* | 24 | 1.37 | 0.010 | 1.1 | 0.20 | 1 | 5 |
| | | | | | | | |
| **Mercury** [a] | … | 0.387 | 0.215 | 6.784 | 0.055 | ≥1? | ? |
| **Venus** [b] | … | 0.723 | 0.032 | 2.159 | 0.815 | ≥0–2? | ? |
| **Earth** [c] | … | 1.000 | 0.028 | 1.988 | 1.000 | ≥1 | 20–140 |
| **Mars** [d] | … | 1.524 | 0.067 | 4.042 | 0.107 | ≥1? | 10? |

**Notes.** N represents the number of identified representative planet analogues. In addition, $a$ is the semi-major axis, $e$ is eccentricity, $i$ is inclination, $m$ is planet mass, nGI is the number of giant impacts (defined by the collision of an object that is at least 10% as massive as the target body), and tGI is the time of the last giant impact suffered by the planet. All quantities (except for N) represent median values. The standard planetary mass ranges considered to identify planet analogues were 0.03–0.17 (Mercury), 0.4–2.0 (Venus and Earth), and 0.05–0.32 (Mars) M$_\oplus$. We averaged the orbital elements of the real planets the last 100 Myr after integrating their orbits. The truncated disk *T7-10* is based on the results of Lykawka & Ito (2019).

[a] If Mercury acquired its iron core through giant impacts, then nGI should be at least equal to 1.

[b] If the absence of Venusian satellites was caused by giant impacts, then nGI should be at least equal to 2 (Alemi & Stevenson 2006). However, the lack of a magnetic dynamo in Venus may be better explained if the planet did not experience such giant impacts comparable to those that formed the Moon (Jacobson et al. 2017).

[c] Evidence suggests that the timing of the Moon-forming giant impact should be within the interval 20–140 Myr (Section 1).

[d] The Martian surface dichotomy may be explained if the planet suffered one giant impact (Marinova et al. 2008).

# APPENDIX

**Table A1.** Properties of the three- or four-terrestrial-planet analogue systems (stringent classification criteria)

| Disk Model | $n3_{mve}$ | $n3_{vem}$ | n4 | AMD | RMC | OS | Lvf (%) | MVr | VEr | MEr |
|---|---|---|---|---|---|---|---|---|---|---|
| D0 | 3 | 32 | 2 | 0.0045 | 71.4 | 37.4 | 9.4 | 0.08 | 1.11 | 0.10 |
| D1 | 0 | 11 | 0 | 0.0019 | 80.8 | 33.1 | 3.9 | … | 0.96 | 0.21 |
| D2 | 1 | 8 | 0 | 0.0040 | 72.8 | 39.3 | 20.8 | 0.13 | 0.96 | 0.13 |
| D3 | 0 | 12 | 0 | 0.0034 | 75.8 | 37.1 | 11.5 | … | 1.29 | 0.10 |
| D4 | 2 | 6 | 0 | 0.0031 | 74.9 | 36.7 | 1.6 | 0.08 | 1.09 | 0.18 |
| D5 | 0 | 2 | 0 | 0.0041 | 75.7 | 39.1 | 25.9 | … | 0.99 | 0.07 |
| **Solar system** | … | … | … | **0.0018** | **89.7** | **37.7** | **0.1–1** | **0.067** | **0.815** | **0.107** |

**Notes.** Please refer to the Table 2 caption for details. The stringent criteria for planetary mass ranges considered to identify planets in analogue systems were 0.03–0.11 (Mercury), 0.6–1.5 (Venus and Earth), and 0.05–0.21 (Mars) $M_\oplus$.

**Table A2.** Representative analogues of Mercury, Venus, Earth, and Mars formed in the three- or four-terrestrial-planet analogue systems (stringent classification criteria)

|  | N | a (au) | e | i (°) | m ($M_\oplus$) | nGI | tGI (Myr) |
|---|---|---|---|---|---|---|---|
| *Mercury* | | | | | | | |
| D0 | 5 | 0.46 | 0.139 | 8.7 | 0.07 | 2 | 27 |
| D1 | 0 | … | … | … | … | … | … |
| D2 | 1 | 0.47 | 0.070 | 11.1 | 0.11 | 2 | 22 |
| D3 | 0 | … | … | … | … | … | … |
| D4 | 2 | 0.45 | 0.153 | 5.7 | 0.07 | 1 | 34 |
| D5 | 0 | … | … | … | … | … | … |
| All disks | 8 | 0.46 | 0.137 | 8.0 | 0.08 | 2 | 27 |
| *Venus* | | | | | | | |
| D0 | 37 | 0.61 | 0.053 | 2.7 | 0.88 | 9 | 8 |
| D1 | 11 | 0.63 | 0.022 | 1.9 | 0.77 | 6 | 10 |
| D2 | 9 | 0.61 | 0.057 | 2.5 | 0.82 | 9 | 10 |
| D3 | 12 | 0.63 | 0.047 | 2.5 | 0.91 | 8 | 12 |
| D4 | 8 | 0.61 | 0.044 | 2.3 | 0.86 | 6.5 | 6 |
| D5 | 2 | 0.63 | 0.048 | 2.3 | 0.79 | 4 | 148 |
| All disks | 79 | 0.62 | 0.050 | 2.5 | 0.87 | 8 | 9 |
| *Earth* | | | | | | | |
| D0 | 37 | 0.98 | 0.058 | 2.9 | 0.79 | 8 | 16 |
| D1 | 11 | 0.96 | 0.020 | 1.8 | 0.80 | 6 | 8 |
| D2 | 9 | 0.95 | 0.070 | 2.6 | 0.86 | 9 | 8 |
| D3 | 12 | 1.01 | 0.041 | 3.0 | 0.71 | 7.5 | 18 |
| D4 | 8 | 0.94 | 0.045 | 2.9 | 0.79 | 6 | 36 |
| D5 | 2 | 0.95 | 0.045 | 2.2 | 0.80 | 6.5 | 11 |
| All disks | 79 | 0.97 | 0.047 | 2.6 | 0.79 | 8 | 16 |
| *Mars* | | | | | | | |
| D0 | 34 | 1.53 | 0.112 | 9.6 | 0.08 | 3 | 21 |
| D1 | 11 | 1.35 | 0.053 | 4.2 | 0.17 | 1 | 1 |
| D2 | 8 | 1.59 | 0.122 | 10.0 | 0.11 | 1.5 | 12 |
| D3 | 12 | 1.49 | 0.141 | 8.7 | 0.07 | 2 | 7 |
| D4 | 6 | 1.44 | 0.105 | 6.3 | 0.14 | 3.5 | 41 |
| D5 | 2 | 1.63 | 0.199 | 10.3 | 0.06 | 1.5 | 111 |
| All disks | 73 | 1.51 | 0.111 | 8.8 | 0.09 | 2 | 11 |
| | | | | | | | |
| **Mercury** | … | 0.387 | 0.215 | 6.784 | 0.055 | ≥1? | ? |
| **Venus** | … | 0.723 | 0.032 | 2.159 | 0.815 | ≥0–2? | ? |
| **Earth** | … | 1.000 | 0.028 | 1.988 | 1.000 | ≥1 | 20–140 |
| **Mars** | … | 1.524 | 0.067 | 4.042 | 0.107 | ≥1? | 10? |

**Notes.** Please refer to the Table 3 caption for details. The stringent criteria for planetary mass ranges considered to identify planet analogues were 0.03–0.11 (Mercury), 0.6–1.5 (Venus and Earth), and 0.05–0.21 (Mars) $M_\oplus$.